\documentclass[12pt]{article}
\pdfoutput=1
\usepackage{amssymb}
\usepackage{amsmath}
\usepackage{bm} 
\usepackage{graphicx}
\usepackage{color}
\usepackage{xcolor}
\usepackage{dsfont}
\usepackage{enumerate}
\usepackage{hyperref}
\usepackage{setspace}
\definecolor{nicered}{rgb}{0.7,0.1,0.1}
\definecolor{nicegreen}{rgb}{0.1,0.5,0.1}
\hypersetup{colorlinks,citecolor= nicegreen,linkcolor= nicered}
\allowdisplaybreaks
\begin{document}
\begin{titlepage}
  \begin{flushright}
    NORDITA-2014-43
  \end{flushright}
  \newcommand{\AddrLiege}{{\sl \small IFPA, Dep. AGO, Universite de
      Liege, Bat B5,\\ \small \sl Sart Tilman B-4000 Liege 1,
      Belgium}}
  \newcommand{\AddrBasel}{\sl \small Department of Physics, University of Basel,
    \\ Klingelbergstr. 82, CH-4056 Basel, Switzerland}
  \vspace*{0.5cm}
\begin{center}
  \textbf{\large
  Reactor mixing angle from hybrid neutrino masses}\\[9mm]
  D. Aristizabal Sierra\footnote{e-mail address:
    {\tt daristizabal@ulg.ac.be}}
  \vspace*{0.4cm}\\
  \AddrLiege.
  \begin{center}
    and
  \end{center}
  I. de Medeiros Varzielas\footnote{e-mail address:
    {\tt ivo.de@unibas.ch}} 
  \vspace*{0.4cm}\\
    \AddrBasel.
\end{center}
\vspace*{0.2cm}
\begin{abstract}
  \onehalfspacing
  In terms of its eigenvector decomposition, the neutrino mass matrix
  (in the basis where the charged lepton mass matrix is diagonal) can
  be understood as originating from a tribimaximal dominant structure
  with small deviations, as demanded by data.  If neutrino masses
  originate from at least two different mechanisms, referred
  to as ``hybrid neutrino masses'', the experimentally observed
  structure naturally emerges provided one mechanism accounts for the
  dominant tribimaximal structure while the other is responsible for the deviations.
  We demonstrate the feasibility of this picture in a fairly
  model-independent way by using lepton-number-violating effective
  operators, whose structure we assume becomes dictated by an
  underlying $A_4$ flavor symmetry. We show that if a second mechanism
  is at work, the requirement of generating a reactor angle within its
  experimental range always fixes the solar and atmospheric angles in
  agreement with data, in contrast to the case where the deviations
  are induced by next-to-leading order effective operators.  We prove 
  this idea is viable by constructing an $A_4$-based
  ultraviolet completion, where the dominant tribimaximal structure
  arises from the type-I seesaw while the subleading contribution is
  determined by either type-II or type-III seesaw driven by a non-trivial $A_4$ singlet
  (minimal hybrid model).
   After finding general criteria, we
  identify all the $\mathbb{Z}_N$ symmetries capable of producing such
  $A_4$-based minimal hybrid models.
\end{abstract}
\end{titlepage}
\setcounter{footnote}{0}
\section{Introduction}
\label{sec:intro}
With the advent of neutrino oscillation data, in particular in the
early 2000's, the tribimaximal (TBM) neutrino mixing pattern
\cite{Harrison:2002er} became a paradigm for model building aiming to
understand the
neutrino mixing through an
underlying flavor symmetry.
The paradigm however, has
been challenged by recent reactor experiments
\cite{An:2012eh,Ahn:2012nd} which have excluded a zero reactor mixing
angle at more than $10\,\sigma$ \cite{Tortola:2012te}.

Expanding the neutrino mass matrix in terms of its eigenvectors, and
with current neutrino data taken into account
\cite{Tortola:2012te,GonzalezGarcia:2012sz,Fogli:2012ua}, one can 
quantify in a useful way
the deviations in the structure of the mass matrix from the
TBM-form
\cite{Sierra:2013ypa}\footnote{As far as we know, this type of eigenvector decomposition was
  first used in the context of a neutrino mass matrix with exact
  TBM-form to show the viability of leptogenesis in models exhibiting
  interplay between type-I and type-II seesaw
  \cite{AristizabalSierra:2011ab}, something which at leading order
  (LO) in the effective operator tower is not possible within the
  type-I seesaw
  \cite{Bertuzzo:2009im,AristizabalSierra:2009ex,Felipe:2009rr}.},
are then small
 due to the ``small'' measured value of the reactor
mixing angle $\theta_{13}$.
Given that the observations are consistent with an approximate TBM-form,
the idea of an underlying
flavor symmetry responsible for the neutrino mixing pattern is arguably well motivated.
Indeed a very
widely explored
possibility is that of a flavor symmetry $G$ provides at the
leading order (LO) a TBM-form neutrino mass matrix and the deviations demanded by
data are obtained through next-to-leading-order (NLO) corrections.
This approach however, at least in the most simple
implementations, does not provide a compelling solution since quite
often a non-zero reactor angle implies either a solar or atmospheric
angles (or both) outside their experimental ranges
\cite{Lin:2009bw, Altarelli:2012bn, Varzielas:2012ai}.

Since the NLO approach is not particularly favored by data, it is important to consider alternative possibilities.
In Ref. \cite{Sierra:2013ypa}, it was argued that the TBM-form plus
small deviations structure, can be interpreted as a hint that
different mechanisms participate in the generation of neutrino masses,
so-called ``hybrid neutrino masses''. The idea is that of
starting with a Lagrangian invariant under $G$, in which after $G$ and
electroweak symmetry breaking one mechanism accounts for the TBM-form
while the deviations are provided by the other mechanism (rather than
by NLO corrections), with both contributions to the neutrino mass
matrix entering at the same order, generically LO.
This type of scenario was investigated before in the context of bi-maximal mixing \cite{Rodejohann:2004cg} and also in the case of TBM \cite{Lindner:2007rs}. 
In \cite{Adulpravitchai:2011rq} it was explored in models with extra dimensions. More recently in \cite{Borah:2013jia, Borah:2013lva, Borah:2014fga},
type-I seesaw \cite{seesaw} was assumed to be
responsible for the leading TBM-form and type-II seesaw
\cite{Schechter:1980gr} contributions introduced the deviations to TBM.

Here we intend to show the feasibility of the whole idea under fairly
general assumptions, in the sense that rather than assuming specific
neutrino mass generating mechanisms we deal with a discussion based on
lepton-number-violating effective operators.  We identify the minimal
$A_4$ representations which render this idea plausible by assuming
that the lepton sector obeys an $A_4$ flavor symmetry similar to the
Altarelli-Feruglio (AF) model \cite{Altarelli:2005yx}, where the
charged lepton Yukawa couplings hierarchies are dictated by a
Froggatt-Nielsen (FN) mechanism \cite{Froggatt:1978nt}.  We
demonstrate that within this context, if the non-zero reactor angle
arises from ``hybrid neutrino masses'' the solar and the atmospheric
angles are always within their experimental ranges, in contrast to the
``standard'' procedure involving NLO effective operators where---in
general---this cannot be guaranteed. More specifically, and in order
to illustrate the general picture, we construct explicitly $A_4$-based
ultraviolet (UV) completions akin to \cite{Varzielas:2010mp}, with the
leading TBM-form originating from type-I seesaw---but importantly the
deviations arise from type-II or type-III
\cite{Foot:1988aq}.
In addition to this and being UV complete with explicit messenger
fields, our construction further differs from that in
\cite{Borah:2013jia}: rather than a flavon $A_4$ triplet with a
certain assumed vacuum expectation value (vev), we use a non-trivial
$A_4$ singlet with a corresponding vev consistently explained by the
alignment sector.

The rest of this letter is organized as follows. In
Sec. \ref{sec:pert-TBM-NMM}, by using effective LO
lepton-number-breaking operators, we study the possible deviations
which can emerge in $A_4$-inspired
frameworks. Sec. \ref{sec:type-ii-seesaw} discusses a viable UV
completion based on type-I plus either type-II or type-III seesaw.
Finally, in
Sec. \ref{sec:concl} we summarize and present our conclusions.
\section{Perturbations of the TBM
  neutrino mass matrix}
\label{sec:pert-TBM-NMM}
If neutrino masses are generated by two different mechanisms, with one
accounting for the TBM structure of the neutrino mass matrix and the
other for the deviations required by experimental data, the full
neutrino mass matrix can be written according to
\begin{equation}
  \label{eq:full-nmm}
  m_\nu = m_\nu^\text{TBM} + \delta m_\nu\ ,
\end{equation}
with the leading contribution $m_\nu^\text{TBM}$ being determined by
\begin{equation}
  \label{eq:mnuTBM}
  m_\nu^\text{TBM}=U_\text{TBM}\cdot \hat m_\nu\cdot U^T_\text{TBM}\ ,
\end{equation}
where $U_\text{TBM}$ corresponds to the leptonic mixing matrix with
mixing angles fixed according to their TBM values
($\sin\theta_{23}=1/\sqrt{2}$,
$\sin\theta_{12}=1/\sqrt{3}$ and
$\sin\theta_{13}=0$). The perturbation matrix $\delta m_\nu$
must produce a non-vanishing and
sizable reactor angle while at the same time yielding not too large
corrections to the other two mixing angles.
At the model-independent
level several conclusions can be drawn, in particular: a non-vanishing
reactor angle requires the perturbation matrix to induce sizable
deviations mainly in the $m_{\nu_{12}}$ and $m_{\nu_{13}}$ entries,
and these deviations in turn imply departures of $\sin\theta_{23}$ and
$\sin\theta_{12}$ from their TBM form, being more pronounced in the
atmospheric sector \cite{Sierra:2013ypa}.

In trying to keep our approach as model-independent as possible,
we develop our analysis by considering first effective $\Delta L=2$ operators accounting for both
$m_\nu^\text{TBM}$ and $\delta m_\nu$.
But to study more in-depth
the viability of this framework 
where deviations from the TBM form arise from a separate mechanism, one must consider particular models
where one may have the structure of $\delta m_\nu$ determined by the
underlying flavor symmetry.
Due to this, we assume a supersymmetric implementation of an $A_4$ lepton flavor symmetry (given its relative simplicity) \cite{Altarelli:2005yx}.

\renewcommand{\arraystretch}{1.3}
\begin{table}
  \centering
  \footnotesize{
    \begin{tabular}{|c||cccc|cc|ccccc|ccc|}
      \hline
      Fields  & $L$ & $e^{c}$ & $\mu^{c}$ & $\tau^{c}$  
      & $H_{d}$ & $H_{u}$ & $\theta$ & $\phi_{l}$ & $\phi_{\nu}$ 
      & $\xi$ & $\tilde{\xi}$  & $\phi_{l}^{0}$ 
      & $\phi_{\nu}^{0}$ & $\xi^{0}$  
      \\\hline\hline
      $A_{4}$ & $\mathbf{3}$  & $\mathbf{1}$
      & $\mathbf{1''}$ & $\mathbf{1'}$  & $\mathbf{1}$  & $\mathbf{1}$
      & $\mathbf{1}$ & $\mathbf{3}$  & $\mathbf{3}$  & $\mathbf{1}$  & $\mathbf{1}$  
      & $\mathbf{3}$  & $\mathbf{3}$  & $\mathbf{1}$
      \\
      $\mathbb{Z}_N$ & $1$  & $-1$ & 
      $-1$ & $-1$ & $0$ & $a$ & $0$ & $0$ & $a'$ & 
      $a'$ & $a'$ & $0$ & $a''$ & $a''$
      \\
      $U(1)_{\textrm{FN}}$ & $0$ & $2$ & $1$ & $0$ & $0$ & $0$
      & $-1$ & $0$ & $0$ & $0$ & $0$ & $0$ & $0$ &  $0$
      \\
      $U(1)_{\textrm{R}}$ & $1$ & $1$ & $1$ & $1$ & $0$ & $0$
      & $0$ & $0$ & $0$ & $0$ & $0$ & $2$ & $2$ &  $2$
      \\
      $U(1)_{\textrm{Y}}$ & $-1/2$ & $+1$ & $+1$ & $+1$ & $-1/2$ 
      & $+1/2$ & $0$ & $0$ & $0$ & $0$ & $0$ & $0$ & $0$  & $0$
      \\
      \hline
    \end{tabular}
    \caption{\it Charge assignments of the different superfields 
      defining the AF model, which we have taken to be responsible
      for the LO TBM-form of the neutrino mass matrix. We have $a'\equiv -2a-2$ and $a''\equiv 4a+4$ and denote $N-1$ as $-1$ for simplicity.   \label{tab:AF-model-CAssign}}
  }

\end{table}
In this context, getting $m_\nu^\text{TBM}$ requires, in addition to
the standard model superfields transforming non-trivially under $A_4$,
the presence of new
superfields
(see Tab. \ref{tab:AF-model-CAssign}, where in contrast to the original model in \cite{Altarelli:2005yx} we have considered $\mathbb{Z}_N$ transformations and
the possibility of the ``up-type'' Higgs transforming non-trivially
under $\mathbb{Z}_N$). $U(1)_Y$ denotes the hypercharge and $U(1)_R$ the R-symmetry,
The so-called alignment superfields (with $U(1)_R$ assignment $2$,
denoted by superscript $0$) will lead to specific non-vanishing vevs
for the so-called flavons (with $U(1)_R$ assignment $0$).

We briefly review the relevant superpotential terms.  In the following, curly brackets denote the contraction of $A_4$ triplets into $A_4$ singlets,
$\left\{ AB \right\} = A_1 B_1 + A_2 B_3 + A_3 B_2$, $\left\{ AB \right\} ''=A_2 B_2 + A_3 B_1 + A_1 B_3$,  $\left\{ AB \right\} '=A_3 B_3 + A_1 B_2 + A_2 B_1$, and similarly for the 3-triplet contractions.

The charged
lepton Yukawa couplings have hierarchical masses due to the
FN mechanism \cite{Froggatt:1978nt}: the
$U(1)_\text{FN}$ flavor symmetry shown
in Tab.~\ref{tab:AF-model-CAssign} is broken by $\langle\theta\rangle$
\begin{equation}
  \label{eq:charged-lepton-Lag}
  W_\ell^\text{Eff}= 
  \frac{y_e}{\Lambda} \lambda^2\, e^c \left\{L \phi_l\right\} \,H_d
  + \frac{y_\mu}{\Lambda}\lambda\,\mu^c\left\{L \phi_l \right\}' \,H_d
  + \frac{y_\tau}{\Lambda}\tau^c \left\{L \phi_l \right\}'' \,H_{d}
  + \text{H.c.}\ ,
\end{equation}
with $\lambda=\langle\theta\rangle/\Lambda$ being the FN suppression
factors and $\Lambda$ the FN cutoff scale.
The superpotential for the neutrinos is
\begin{equation}
  \label{eq:neutral-Lag}
  W_\nu^\text{Eff} = 
  \frac{\left(x_{A}\xi + \tilde{x}_{A}\tilde{\xi} \right)}{\Lambda^2}
  \left\{LL\right\}\,H_u\,H_u
  + \frac{x_{B}}{\Lambda^2}\left\{\phi_{\nu}LL\right\}\,H_u\,H_u\ .
\end{equation}
The LO TBM comes from the flavon vevs:
\begin{equation}
  \label{eq:vev-alignment}
  \langle\phi_l\rangle=v_l(1,0,0)\ ,
  \quad
  \langle\phi_\nu\rangle=v_\nu(1,1,1)\ ,
  \quad
  \langle\xi\rangle=u\ ,
\end{equation}
provided by the alignment superpotential:
\begin{align}
  \label{eq:scalar-potential}
  W_\phi& = M\left\{ \phi^{0}_{l}\phi_{l} \right\}
  + g\left\{ \phi_{l}^{0}\phi_{l}\phi_{l} \right\} 
  + g_{1}\left\{ \phi_{\nu}^{0}\phi_{\nu}\phi_{\nu} \right\}
  + g_{2}\tilde{\xi}\left\{ \phi_{\nu}^{0}\phi_{\nu} \right\}
  + g_{3}\xi^{0}\left\{ \phi_{\nu}\phi_{\nu} \right\}
  \nonumber\\
  &+ g_{4}\xi^{0}\xi\xi
  + g_{5}\xi^{0}\xi\tilde{\xi}
  + g_{6}\xi^{0}\tilde{\xi}\tilde{\xi} \ .
\end{align}

These superpotential terms
are invariant under all the symmetries in Table \ref{tab:AF-model-CAssign}. 
The presence of the
$\mathbb{Z}_N$ symmetry implies that charged lepton and neutrino
masses arise from two independent sets of fields, while the $U(1)_R$
separates the superfields as those that correspond to fermions of the
standard model (first four columns in
Tab. \ref{tab:AF-model-CAssign}), alignment (or ``driving'')
fields (last three columns in Tab. \ref{tab:AF-model-CAssign}), and symmetry breaking fields.

After electroweak and flavor symmetry breaking the set of
interactions in (\ref{eq:neutral-Lag}), combined with the vacuum
alignment in (\ref{eq:vev-alignment}) and the appropriate $A_4$ group
index contractions, yields for the effective neutrino mass matrix
\begin{equation}
  \label{eq:Dirac-RHN-MM}
  m_\nu^\text{TBM}=
  \begin{pmatrix}
    A+2B/3 & -B/3  & -B/3\\
    \cdot  & 2B/3  & A-B/3\\
    \cdot  & \cdot & 2B/3
  \end{pmatrix}\ ,
\end{equation}
with $\langle H_u\rangle=v_u$ and the parameters in the RH neutrino
mass defined as
\begin{equation}
  \label{eq:A-B-parameters}
  A=2\,x_A\,v_u^2\,\frac{u}{\Lambda^2} \qquad\mbox{and}\qquad
  B=2\,x_B\,v_u^2\,\frac{v_\nu}{\Lambda^2}\ .
\end{equation}
In general this structure is maintained only at LO, unless a specific
UV completion guarantees the absence of NLO contributions
\cite{Varzielas:2010mp}.
The NLO contributions modify $m_\nu^\text{TBM}$ directly through a
change in the neutrino terms adding to (\ref{eq:neutral-Lag}) or
indirectly by shifting the vacuum alignment
in~(\ref{eq:vev-alignment}).  Constraining the discussion to only this
setup, this implies that ${\cal O}(\delta m_\nu/m_\nu^\text{TBM})\sim
\text{vev}/\Lambda\equiv \delta$. For values of $\delta$ of ${\cal
  O}(0.1)$, these corrections can account for a non-vanishing reactor
angle but at the same time imply deviations of the solar and
atmospheric angle from their TBM values, often outside of their
experimental ranges \cite{Lin:2009bw, Altarelli:2012bn, Varzielas:2012ai}.
For smaller $\delta$, below $\sim 0.1$, the TBM structure is less affected.

If additional ($A_4$) flavons are present, they can allow additional
lepton-number-breaking effective operators that deviate from
$m_\nu^\text{TBM}$. If the new operators are less suppressed than the
the NLO operators of the ``original'' model, the deviations from TBM
will be {\it naturally} dictated by the new contribution, which we
denote as $\delta m_\nu^\text{Extra}$. This can happen if
$\delta m_\nu^\text{NLO}<\delta m_\nu^\text{Extra}\sim{\cal O}(0.1)$, e.g. if the NLO operators are
forbidden by the UV completion. Then $\delta
m_\nu^\text{Extra}$ must generate deviations consistent with
data.
Given the lepton doublets transformation properties, three $A_4$
assignments for an extra flavon are possible: a triplet $\mathbf{3}$
\cite{Borah:2013jia}, a singlet
$\mathbf{1''}$ or a singlet $\mathbf{1'}$. For the triplet to produce
a deviation in the TBM leading structure, the flavon vev should differ
from that of $\phi_\nu$ in (\ref{eq:vev-alignment}), otherwise
$\delta m_\nu^\text{Extra}$ will just shift the parameter $B$ in
(\ref{eq:Dirac-RHN-MM}), keeping its TBM-form \footnote{The same will
  happen with the parameter $A$ if the new flavon is a trivial $A_4$
  singlet, $\mathbf{1}$.}. Instead, with the new flavon transforming
as $\mathbf{1''}$ or $\mathbf{1'}$, a non-vanishing vev will be
sufficient to produce deviations from TBM, so the non-trivial singlet
choices are simpler.

A minimal hybrid singlet-based AF-inspired model will then consist of the AF fields
plus new flavons, namely $\xi'$ and its corresponding ``driving'' field
$\xi^{0'}$ or $\xi''$ and $\xi^{0''}$. Two conditions have to be
fulfilled for the emerging scheme to be phenomenologically
consistent: $(a)$ the TBM deviations induced by the new Yukawa
operators arising from the presence of the singlet should be
consistent with data
\cite{Tortola:2012te,GonzalezGarcia:2012sz,Fogli:2012ua}; $(b)$ the
new ``driving sector'' operators should lead to a non-vanishing
singlet vev and at the same time not spoil the vev alignments (\ref{eq:vev-alignment}) assuring
the leading order TBM structure.

We consider separately the two types of minimal $A_4$-based hybrid models:
with $\xi''$ or with $\xi'$. The respective dimension six effective operators would be:
\begin{equation}
  \label{eq:hybrid-dim6-effective-operators}
  {\cal O}_6 =\frac{z}{\Lambda^2}\{L\,L\}'\,\xi''\,H_u\,H_u\ ,
  \qquad
  {\cal O}_6 =\frac{z}{\Lambda^2}\{L\,L\}''\,\xi'\,H_u\,H_u\ .
\end{equation}
$\mathbb{Z}_N$ invariance then fixes the charge of the new
flavon to match that of the AF trivial singlet flavon $\xi$.
Such a charge assignment typically renders the hybrid scheme ineffective due to unavoidable additional invariants. Specifically,
if some $\left\{X\right\}\xi$ coupling exists ($X$ being a
combination of fields belonging to the UV completed model, particularly $A_4$ invariants made of two triplets),
$\mathbb{Z}_N$ invariance will also allow the coupling $\left\{X\right\}'\xi''$ (or
$\left\{X\right\}''\xi'$).
In other
words, the $\mathbb{Z}_N$ equality implies that both operators,
the one yielding TBM and the one responsible for the deviations, will
stem from the same (UV complete) mechanism.
This is certainly the case in the AF-inspired models,
where equal $\mathbb{Z}_N$ charges lead the $\xi$ and $\xi''$ (or $\xi'$) to contribute through type-I seesaw.
The conclusion is more general though: minimal hybridization
will often require a distinct charge under some Abelian symmetry, even though the assignments under the non-Abelian symmetry are already distinct. We will refer to this as the
{\it hybridization statement}.

There is another argument supporting an even stronger {\it
  hybridization statement}, and it has to do with the hybridization
condition $(b)$. In a specific UV completion of the AF-inspired models, a mismatch between
$\xi$ and $\xi''$ (or $\xi'$) charges can be given by a sign mismatch (where a charge of $-a$ corresponds to $N-a$ under $\mathbb{Z}_N$). Such a mismatch avoids the coupling $\left\{X\right\}'\xi''$ (or $\left\{X\right\}'' \xi'$).
However, new terms at the ``driving sector'' level will be
allowed, namely
\begin{equation}
  \label{eq:new-driving-SP}
  W^\text{New}_\phi\supset \xi''\{\phi_l^0\phi_\nu\}' + 
  \xi''\{\phi_\nu^0\phi_l\}'\ ,
\end{equation}
or the equivalent terms for $\xi'$. The first term destroys the $\langle\phi_l\rangle$ alignment in
(\ref{eq:vev-alignment}), while the second term spoils the
$\langle\phi_\nu\rangle$ alignment, and therefore any of them harming
the LO TBM structure. A mismatch beyond just a sign flip is thus
mandatory in the AF-inspired models. 
This constraint applies in general to models where some flavons or alignment fields are neutral under the Abelian symmetries.

At the effective level, such a
mismatch is possible only if $\xi$ and $\xi''$ (or $\xi'$)
participate in operators of different dimensionality.
If one aims to argue that the hybrid structure accounts for
the small TBM deviations it is
desirable to have the deviations appear at higher order, the simplest cases are then
\begin{align}
  \label{eq:type-II-contributions-singlet-dp}
  {\cal O}_7&=\frac{z}{\Lambda^3}\{L\,L\}'\,\xi'\xi'\,H_u\,H_u
  \qquad
  \stackrel{\text{After SB}}{\longrightarrow}
  \qquad
  \delta m_\nu= \delta m^{33}_{12}=\epsilon
  \begin{pmatrix}
    0     & 1     &  0\\
    1     & 0     &  0\\
    0     & 0     &  1
  \end{pmatrix}\ ,
  \\
  \label{eq:type-II-contributions-singlet-p}
  {\cal O}_7&=\frac{z}{\Lambda^3}\{L\,L\}''\,\xi''\xi''\,H_u\,H_u
  \qquad \stackrel{\text{After SB}}{\longrightarrow} \qquad \delta
  m_\nu=\delta m^{22}_{13}=\epsilon
  \begin{pmatrix}
    0 & 0 & 1\\
    0 & 1 & 0\\
    1 & 0 & 0
  \end{pmatrix}\ ,
\end{align}
where we have defined $\epsilon\equiv z\,v_u^2\,\bar u^2/\Lambda^3$
and have assumed that alignment superpotential leads to $\langle\xi^{''(')}\rangle=\bar u\neq 0$. Note that with the
deviations dictated by ${\cal O}_7$ the $\mathbb{Z}_N$ charge of the
non-trivial singlet flavon is $-1-a$, half the charge of $\xi$. 

In summary, if the leading TBM-form of the neutrino mass matrix arises
from this type of $A_4$ setup, the simplest lepton-number-breaking
operators which can induce deviations from the TBM-form are determined
by the presence of a flavon transforming as a non-trivial $A_4$
singlet \cite{Varzielas:2012ai} (see also \cite{Brahmachari:2008fn, Barry:2010zk, Shimizu:2011xg}).
Thus, within this context, two minimal setups can be defined, namely
\begin{align}
  \label{eq:setups}
  A_4-\Xi''\;\text{model}:&\qquad m_\nu=m_\nu^\text{TBM} + \delta m^{33}_{12}
  \qquad\mbox{with}\;\;\Xi''\equiv\xi'\xi'\ ,
  \\
  A_4-\Xi'\;\text{model}:&\qquad m_\nu=m_\nu^\text{TBM} + \delta m^{22}_{13}
  \qquad\mbox{with}\;\;\Xi'\equiv\xi''\xi''\ .
\end{align}
Assuming $v_\nu\sim u$ and $x_A\sim x_B$, the relative size between
the two terms becomes determined by $\delta
m_\nu^\text{Extra}/m_\nu^\text{TBM}=z\bar u^2/2x_Au\Lambda$, with the
precise numerical ranges fixed by the requirement of having a mixing
pattern compatible with neutrino oscillation data
\cite{Tortola:2012te,GonzalezGarcia:2012sz,Fogli:2012ua}.

\begin{figure}
  \centering
  \includegraphics[scale=0.62]{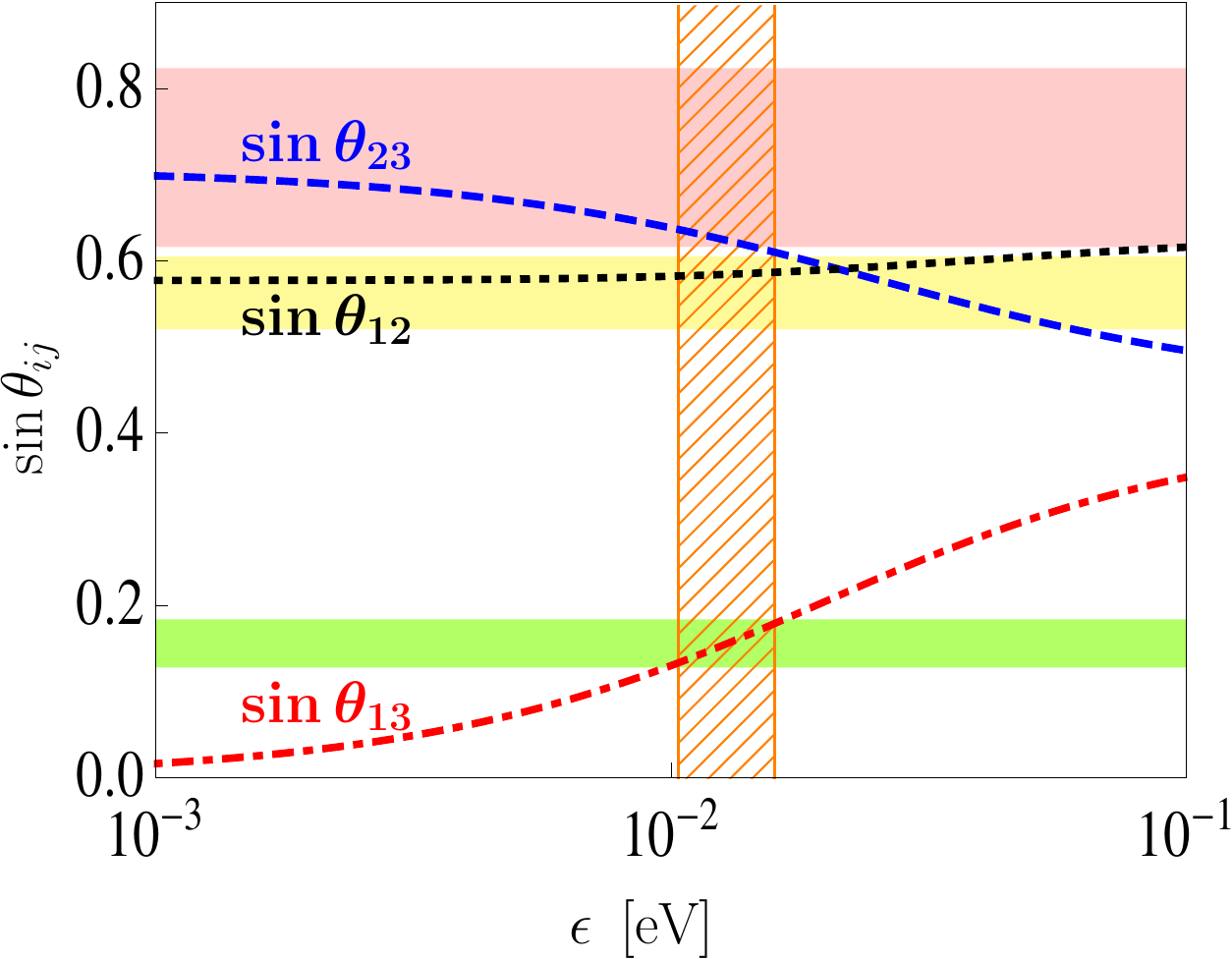}
  \includegraphics[scale=0.62]{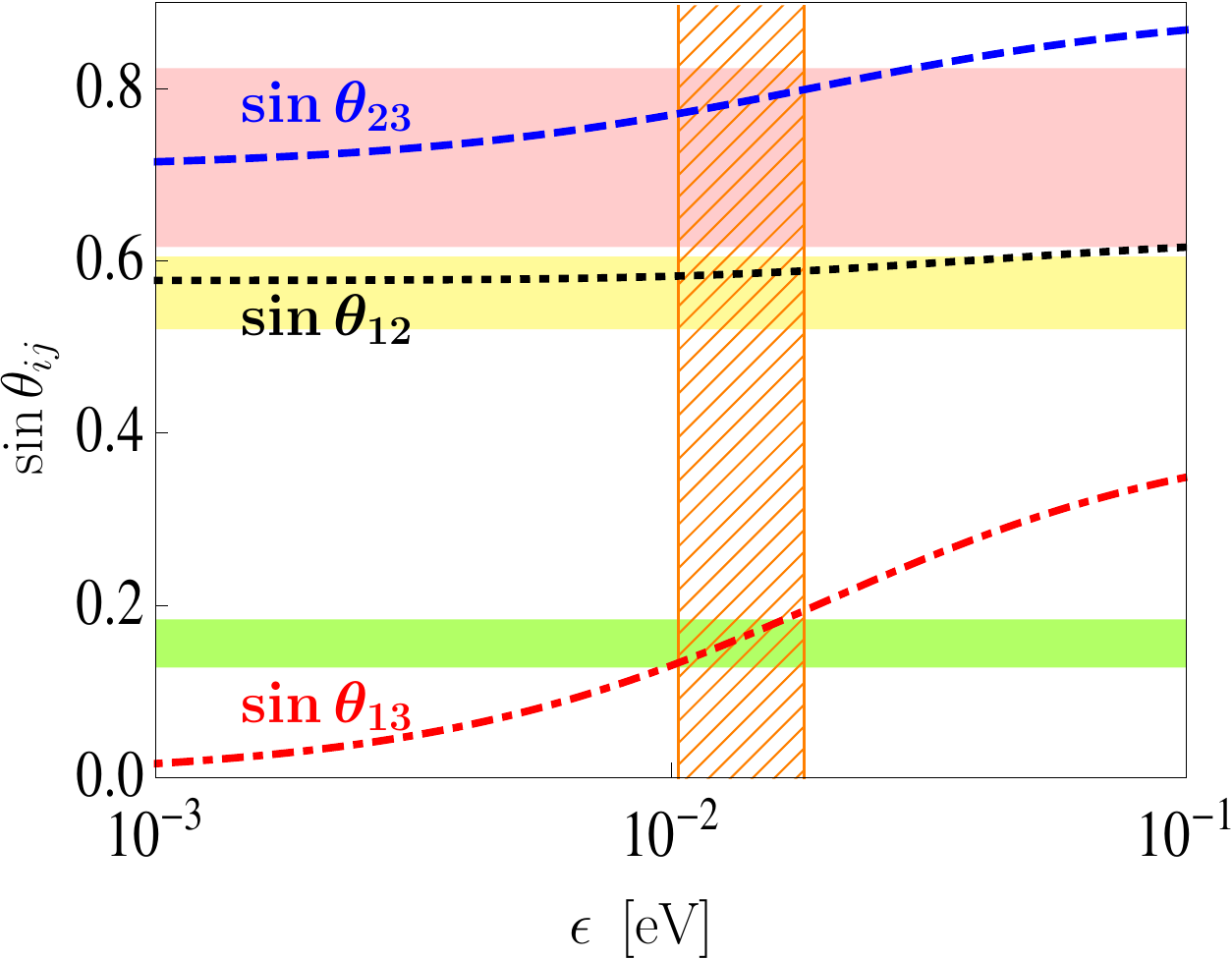}
  \caption{\it Deviations on the different mixing angles induced by
    the perturbation matrices $\delta m_{12}^{33}$ (left plot) and
    $\delta m_{13}^{22}$ (right plot). The horizontal stripes indicate
    the $3\sigma$ allowed range for the atmospheric angle (top
    stripe), solar angle (middle stripe) and reactor angle
    (bottom stripe). The hatched vertical stripe
    indicates the range that parameter $\epsilon$
    should have in order for the deviations to be consistent with
    current neutrino data
    \cite{Tortola:2012te,GonzalezGarcia:2012sz,Fogli:2012ua}.   \label{fig:perturbations}}
\end{figure}
Taking $m_\nu^\text{TBM}$ as given in (\ref{eq:mnuTBM}), and
fixing the lightest light neutrino mass to be somewhere below
$10^{-2}\,$~eV the full neutrino mass matrix becomes a function of a
single parameter, $\epsilon$, i.e. $m_\nu=m_\nu(\epsilon)$. Thus, in
order to check for the viability of the resulting models, one can
diagonalize $m_\nu(\epsilon)$ and see whether for the $\epsilon$ range
where $\sin\theta_{13}$ falls into its measured value the remaining
two mixing angles are also within their experimental
ranges. Fig. \ref{fig:perturbations} shows the results in both cases
for the normal mass spectrum, left plot for the
$A_4-\Xi''\;\text{model}$ while the right plot for the
$A_4-\Xi'\;\text{model}$. We fixed $m_{\nu_1}$ according to
$10^{-3}\,$~eV. These results prove that as long as $\epsilon\subset
[1.2,2.2]\times 10^{-2}\,$~eV, which implies $\delta
m_\nu^\text{Extra}/m_\nu^\text{TBM}\sim \epsilon/m_{\nu_3}\sim 0.4$,
one is always able to obtain a mixing pattern consistent with
data. More importantly,
{\it for the range where the reactor angle
  falls in its $3\sigma$ experimental range, the atmospheric and solar
  angles are also in the $3\sigma$ range}.
  This follows directly from implementing hybrid seesaw in the particular case of $A_4$ with non-trivial singlets,
   and compares favorably with
the AF model, where NLO contributions required to match the observed reactor angle do not necessarily imply that the
solar and atmospheric angles are in agreement with data
\cite{Lin:2009bw, Altarelli:2012bn} (this is also the case for UV completions \cite{Varzielas:2012ai}).

\subsection{Sum-rules and constraints on neutrino observables}
\label{sec:sum-rules}
The complex eigenvalues of the leading order TBM mass matrix in
(\ref{eq:Dirac-RHN-MM}) are given by $A$, $B-A$ and $B+A$. They can be
identified with the TBM eigenvectors
$\boldsymbol{v_1}=(-2,1,1)/\sqrt{6}$,
$\boldsymbol{v_2}=(1,1,1)/\sqrt{3}$, and
$\boldsymbol{v_3}=(0,-1,1)/\sqrt{2}$, thus implying: 
\begin{equation}
  \label{eq:eigenvalues}
  m_{\nu_1}=A+B\ ,\quad
  m_{\nu_2}=A\ , \quad m_{\nu_3}=B-A\ .
\end{equation}
Therefore, at the TBM leading order level neutrino masses obey the
following sum-rule:
\begin{equation}
  \label{eq:sum-rule}
  m_{\nu_1} = 2m_{\nu_2} +  m_{\nu_3}\ .
\end{equation}
\renewcommand{\arraystretch}{1.4}
\begin{table}[t!]
  \centering
  \footnotesize{
  \begin{tabular}{|c|c|c|c|c|c|c|c|c|}\hline
    {\sc Obs.}& \multicolumn{2}{|c|}{$|m_\text{Light}|$ [eV]}
    & \multicolumn{2}{|c|}{$\sum m_{\nu_i}$ [eV]}
    & \multicolumn{2}{|c|}{$m_\beta$ [eV]}
    & \multicolumn{2}{|c|}{$\langle m_{ee}\rangle$ [eV]}\\\hline
    {\sc Spect.} & NO ($|m_1|$) & NO ($|m_3|$)
    & NO           & IO
    & NO           & IO
    & NO           & IO\\\hline\hline
    $\varepsilon=0$ 
    & $\gtrsim 0.0014$ & $\bullet$
    & $\gtrsim0.078$   & $\bullet$
    & $\gtrsim0.015$   & $\bullet$
    & $\gtrsim0.0031$  & $\bullet$\\\hline
    $\varepsilon=0.1$
    & $\gtrsim 0.012$ & $\gtrsim 0.062$
    & $\gtrsim0.074$   & $\gtrsim 0.22$
    & $\gtrsim0.014$   & $\gtrsim 0.077$
    & $\gtrsim0.0029$  & $\gtrsim 0.054$\\\hline
    $\varepsilon=0.2$
    & $\gtrsim 0.0098$ & $\gtrsim 0.034$
    & $\gtrsim0.070$   & $\gtrsim 0.15$
    & $\gtrsim0.012$   & $\gtrsim 0.057$
    & $\gtrsim0.0023$  & $\gtrsim 0.054$\\\hline
    $\varepsilon=0.3$
    & $\gtrsim 0.0076$ & $\gtrsim 0.019$
    & $\gtrsim0.065$   & $\gtrsim 0.12$
    & $\gtrsim0.010$   & $\gtrsim 0.049$
    & $\gtrsim0.0013$  & $\gtrsim 0.047$\\\hline
  \end{tabular}}
\caption{\it Lower limits on neutrino observables for 
  the normal (NO) and inverted (IO) ordering of the masses. The dot indicates that the IO is not 
  allowed in the exact TBM limit. The results have 
  been taken from \cite{Barry:2010yk}.}
  \label{tab:neutrino-obs}
\end{table}
As has been discussed at length in \cite{Barry:2010yk}, neutrino
mass sum-rules add further constraints on neutrino observables: the
lightest neutrino mass ($m_\text{light}$), the sum of absolute
neutrino masses ($\sum m_{\nu_i}$), the kinematic electron neutrino
mass in $\beta$ decay ($m_\beta$) and the effective mass for
$0\nu\beta\beta$ ($\langle m_{ee}\rangle$). In the presence of
deviations from the TBM structure the sum-rules and
corresponding constraints change. For deviations given by
(\ref{eq:type-II-contributions-singlet-dp}) and
(\ref{eq:type-II-contributions-singlet-p}), the eigenvalues---at order
$\epsilon$---are instead given by
\begin{equation}
  \label{eq:eigenvalues-mnu-plus-pert}
  m_{\nu_1}=A + B - \frac{\epsilon}{2}\ ,
  \quad
  m_{\nu_3}=A + \epsilon\ ,
  \quad
  m_{\nu_3}=B - A + \frac{\epsilon}{2}\ ,
\end{equation}
which in turn imply the following modified sum-rule
\begin{align}
  \label{eq:sum-rules-pert-xi-dp}
  2 m_{\nu_2} + m_{\nu_3} - m_{\nu_1}= 3\epsilon\ .
\end{align}
In \cite{Barry:2010yk} this type of deviation from the sum rule was parametrized as
\begin{align}
  \label{eq:sum-rules-pert-xi-dp-ref}
  2 m_{\nu_2} + m_{\nu_3} - m_{\nu_1}=
  \varepsilon\,|m^0_\text{heaviest}|\,e^{i\phi_3}\ .
\end{align}
This means that the results in \cite{Barry:2010yk} can be directly
translated to our case through the equality
$\varepsilon=3|\epsilon|/|m^0_\text{heaviest}|$. With ${\cal
  O}(\epsilon)\sim 0.01\,$~eV, as required by neutrino mixing data
(see Fig.  \ref{fig:perturbations}, hatched (orange) vertical stripe),
and fixing the heaviest neutrino mass to be ${\cal
  O}(|m_\text{heaviest}|)\sim 0.1\,$~eV one can estimate $\varepsilon$
to be order $0.1-0.3$. Thus, once the hybrid contribution is added the
lower limits on the neutrino observables will obey the constraints
found in \cite{Barry:2010yk} for $\varepsilon=0.1-0.3$, which---for
completeness---we summarize in Tab. \ref{tab:neutrino-obs}. Note that
in the exact TBM limit the inverted spectrum is not allowed, but
becomes possible once the deviations are added.


\section{Renormalizable $A_4$ hybrid seesaw
\label{sec:type-ii-seesaw}}
Given the viability of generating a non-vanishing reactor
mixing angle from two different sets of $\Delta L=2$ effective
operators, it is interesting to discuss UV complete
realizations.  For that aim we will have type-I seesaw
responsible for the leading TBM-form while $\delta m_\nu$ arises from type-II (assumed from now up to Sec. \ref{sec:type-iii}) or type-III seesaw (assumed in Sec. \ref{sec:type-iii}). We will specialize to the case of
the $A_4-\Xi'$ model, results for the $A_4-\Xi''$ model follow
directly from what will be discussed below.

Getting $m_\nu^\text{TBM}$ from type-I seesaw requires the presence of
RH neutrinos superfields ($\nu^c$), which we take to transform as
$A_4$ triplets, $\mathbf{3}$. The seesaw Yukawa couplings
$L\,\nu^c\,H_u$ require the RH neutrinos to have a $\mathbb{Z}_N$
charge equal to $-1-a$, and this requirement in turn implies that the
neutral sector flavons $\phi_\nu$, $\xi$ and $\tilde\xi$ should have
$\mathbb{Z}_N$ charges equal to $2+2a$, rather than as $-2-2a$ as they
do when effective operators are used instead. With $\phi_\nu$, $\xi$
and $\tilde\xi$ having this charge, the scalar potential terms in
(\ref{eq:scalar-potential}) require the neutral sector driving fields
to have charges equal to $-4-4a$. All in all, the introduction of the
type-I seesaw changes the transformation properties of the neutral
sector flavon fields in Tab. \ref{tab:AF-model-CAssign}, while leaving
the charged lepton sector fields unchanged.

\begin{figure}
  \centering
  \includegraphics[scale=0.8]{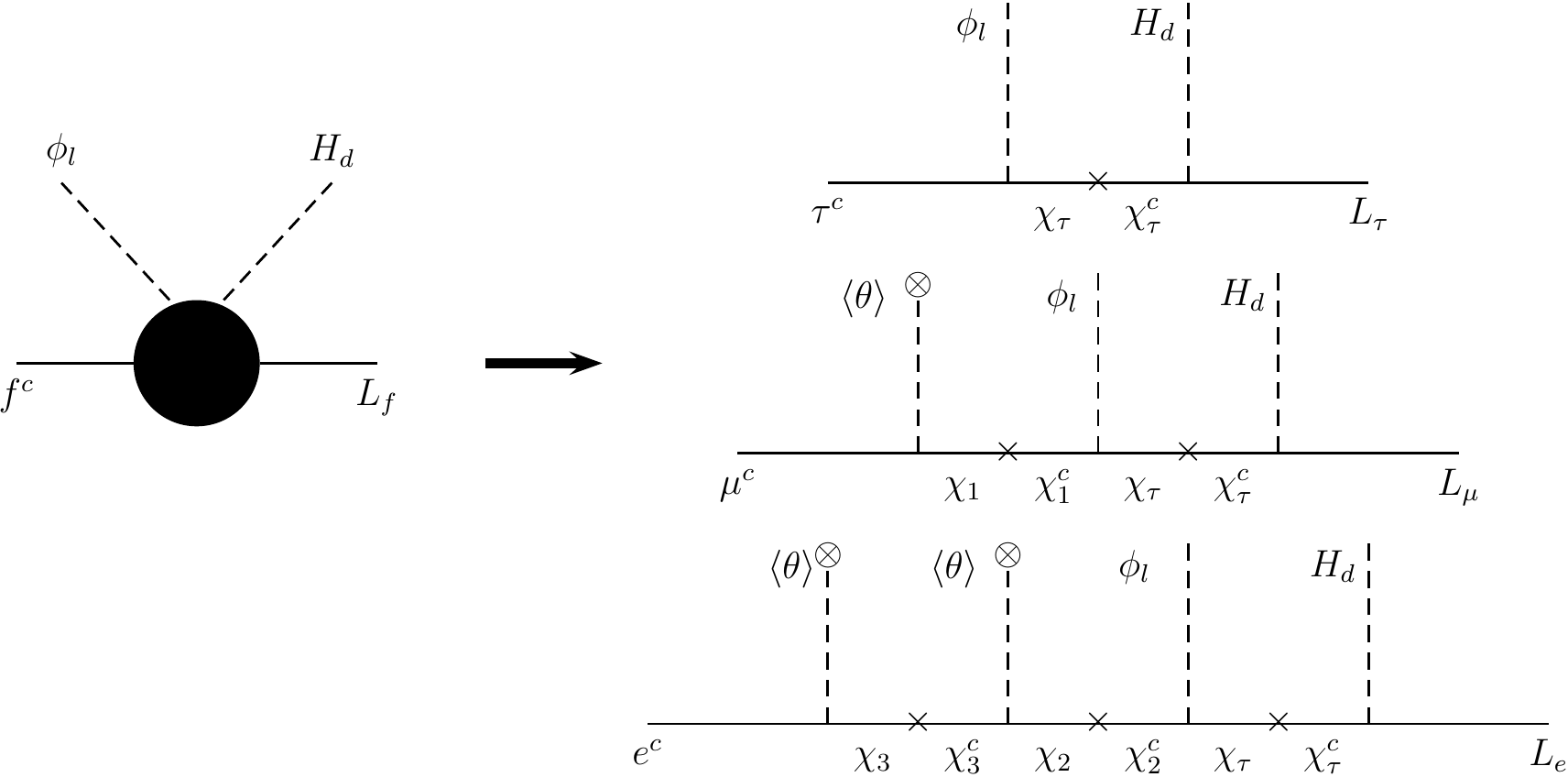}
  \caption{\it Effective charged lepton Yukawa operators arising from
    integrating out the heavy vector-like fields $\chi_A$ (messenger
    fields) entering in the set of renormalizable interactions in the
    superpotential in (\ref{eq:charged-lepton-Lag-renormalizable}).}
  \label{fig:effective-CL-operators}
\end{figure}

\renewcommand{\arraystretch}{1.3}
\begin{table}
  \centering
  \begin{tabular}{|c||cccc|cccc|}
    \hline
    Fields & $\chi_{\tau}$ & $\chi_{1}$ & $\chi_{2}$ & $\chi_{3}$
    & $\chi_{\tau}^{c}$ & $\chi_{1}^{c}$ & $\chi_{2}^{c}$ & $\chi_{3}^{c}$
    \\\hline\hline
    $A_{4}$ & $\mathbf{3}$  & $\mathbf{1'}$  & $\mathbf{1}$ &
    $\mathbf{1}$  & $\mathbf{3}$  & $\mathbf{1''}$ & $\mathbf{1}$
    & $\mathbf{1}$
    \\
    $\mathbb{Z}_{N}$ & 1  & 1  & 1  & 1  &
    $-1$  & $-1$ & $-1$ & $-1$
    \\
    $U(1)_{\textrm{FN}}$ & $0$ & $0$ & $0$ & $-1$ & $0$ & $0$ & $0$ 
    & $+1$
    \\
    $U(1)_{\textrm{R}}$ & $1$ & $1$ & $1$ & $1$ & $1$ & $1$ & $1$
    & $1$
    \\
    $U(1)_{\textrm{Y}}$ & $-1$ & $-1$ & $-1$ & $-1$ & $+1$ & $+1$
    & $+1$ & $+1$
    \\
    \hline
  \end{tabular}
  \caption{\it Messenger sector for the charged lepton sector.
    The $\chi$ fields 
    are the FN messenger superfields responsible for the charged lepton Yukawa 
    couplings hierarchies.   \label{tab:messenger-fields}}

\end{table}
Since we are considering a UV completion, we have only renormalizable
interactions, the charged lepton superpotential in
(\ref{eq:charged-lepton-Lag}) will be obtained through messenger
superfields transforming non-trivially under $\mathbb{Z}_N$ and $U(1)_Y$
(see Tab. \ref{tab:messenger-fields}). Apart from the changes to $\mathbb{Z}_N$ (matching what is required due to the neutrino sector), these are the same messengers that were employed originally in \cite{Varzielas:2010mp}, meaning
we also adopt the convention of
coupling the left-handed (RH) leptons to $H_d$ ($\theta$), in which
case the explicit renormalizable superpotential reads
\cite{Varzielas:2010mp}:
\begin{align}
  \label{eq:charged-lepton-Lag-renormalizable}
  W_\ell&= y_{L\tau}\,\{L\,\chi^c_\tau\}\,H_d
  + y_{\tau\tau}\,\tau^c\,\{\chi_\tau\,\phi_l\}
  + y_{1\tau}\,\chi_1^c\,\{\chi_\tau\,\phi_l\}'
  + y_{\mu 1}\,\mu^c\,\chi_1\,\theta
  + y_{2\tau}\,\chi_2^c\,\{\chi_\tau\,\phi_l\}
  \nonumber\\
  & + y_{32}\,\chi_3^c\,\chi_2\,\theta
  + y_{e3}\,e^c\,\chi_3\,\theta
  + M_{\chi_A}\,\chi_A\,\chi_A
  + \text{H.c.}\ ,
\end{align}
with the different couplings $y_{ij}$ being order-one numbers and the
$A$ subscript denoting the different messengers in their respective
mass terms.  Integrating out the messengers yields the effective
superpotential in~(\ref{eq:charged-lepton-Lag}), as shown in
Fig. \ref{fig:effective-CL-operators}.

The neutral renormalizable superpotential consist of two pieces, one
accounting for the type-I seesaw interactions and a second piece
responsible for the other sector:
\begin{equation}
  \label{eq:type-I-and-type-II-Lag}
  W_\nu = W_\nu^{(I)} + W_\nu^{(II)}\ .
\end{equation}
With the $\mathbb{Z}_N$ assignments, the type-I seesaw superpotential can then be written as
\begin{equation}
 \label{eq:type-I-seesaw-Lagrangian}
 W_\nu^{(I)}= y_N\,\left\{ L\,\nu^{c} \right\}\,H_u + 
\left(x_{A}\,\xi + \tilde{x}_{A}\,\tilde{\xi} \right)
\left\{\nu^{c}\,\nu^{c} \right\} 
 + x_{B}\left\{\phi_{\nu}\,\nu^{c}\,\nu^{c}\right\}\ ,
\end{equation}
with the corresponding alignment terms ensuring the appropriate
flavon vevs given in (\ref{eq:scalar-potential}).
After flavor and electroweak symmetry breaking, the Dirac mass matrix and RH
neutrino mass matrix
combine through
type-I seesaw into a TBM-form for the effective neutrino mass matrix,
$m_\nu^\text{TBM}=m_D\left(M_N\right)^{-1}m_D^T$.

Before specifying the type-II sector we will discuss some generalities
regarding the flavon field $\xi'$. As has been pointed out in
Sec. \ref{sec:pert-TBM-NMM}, its $\mathbb{Z}_N$ charge is fixed to be
$-1-a$ in the effective approach. As we will show, through a specific UV completion there is also the possibility to allow the charge to be the opposite choice, so we have two choices:
$\pm(1+a)$. Depending on the $a$ charges and the cyclic group order
$N$, the presence of $\xi'$ and the corresponding ``driving'' flavon
$\xi^{0'}$ might allow the construction of the following relevant
``driving sector'' operators:
\begin{equation}
  \label{eq:superpotential-driving-sector-model}
  W^\text{New}_\phi= \bar g_1\xi'\{\phi^0_l\phi_l\}'' 
  + \bar g_2 \xi'\{\phi_\nu^0\phi_l\}''
  + \bar g_3 \xi'\{\phi_l^0\phi_\nu\}''
  + \bar g_4 \xi^{0'}\{\phi_l\phi_\nu\}''
  + \bar g_5 \xi^{0'}\xi'\xi'\ .
\end{equation}
Regardless of the choice the first three operators should be forbidden
otherwise the LO TBM structure dictated by the type-I seesaw sector
will be spoiled. The last two terms instead will be the ones
accounting for a non-vanishing $\langle\xi'\rangle$, and so their
presence is desirable. Since the positive charge choice, $1+a$,
for $\xi'$ automatically guarantees the presence of both, provided the
driving field $\xi^{0'}$ charge is $-2(1+a)$, we regard this charge
assignment as the most compelling one. Note that by doing this there will be some subtlety in determining the correct $A_4$ contraction of the $L$, which determine the shape of the mass matrix contribution and whether the respective model is a $\Xi'$ or a $\Xi''$ model.

\begin{figure}
  \centering
  \includegraphics[scale=0.75]{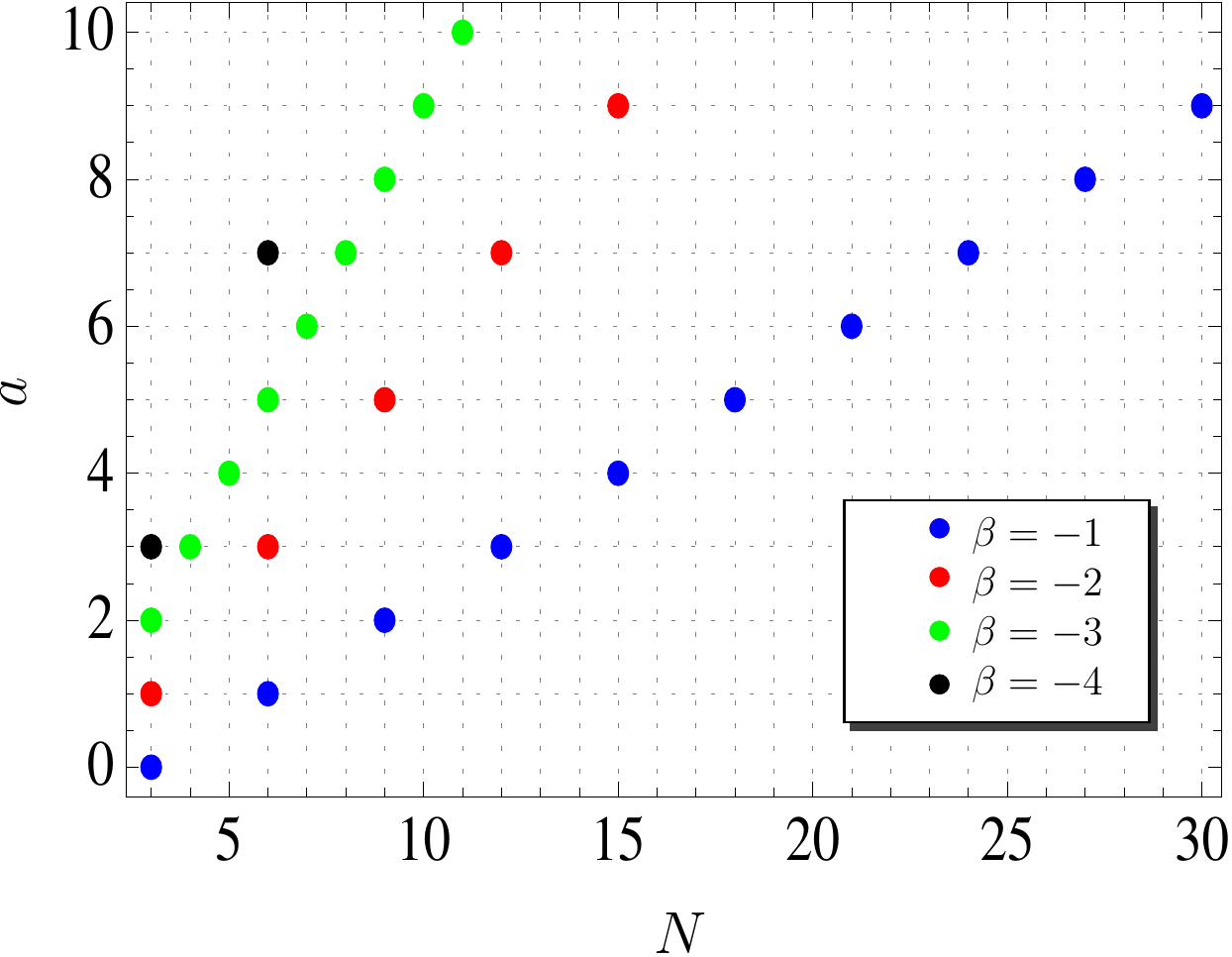}
  \caption{\it Group order $N$ and possible charge assignments for the
    charge $a$ which allow the presence of all or part of the
    dangerous operators. The points in the grid with no colored mark
    correspond to the viable models. No
    viable models exist for $N=3$.}
  \label{fig:n-and-a-assignments-allowing-dangerous-T}
\end{figure}

The allowed $a$ and $N$ choices, i.e. the choices for which the first
three terms in (\ref{eq:superpotential-driving-sector-model}) are
forbidden can be determined in complete generality by considering
their $\mathbb{Z}_N$ transformation properties.  Taking
$\eta_N \equiv e^{i 2\pi /N}$, the dangerous operators in
(\ref{eq:superpotential-driving-sector-model}) transform according to
\begin{align}
  \label{eq:ZN-trans-dangerous}
  \xi'\{\phi^0_l\phi_l\}''&\to\eta_N^{(1+a)}\xi'\;\{\phi^0_l\phi_l\}''\ ,\\
  \xi'\{\phi_\nu^0\phi_l\}''&\to\eta_N^{-3(1+a)}\xi'\;\{\phi_\nu^0\phi_l\}''\ ,\\
  \xi'\{\phi_l^0\phi_\nu\}''&\to\eta_N^{3(1+a)}\xi'\;\{\phi_l^0\phi_\nu\}''\ .
\end{align}
Thus, these terms will be allowed provided the following conditions hold
\begin{equation}
  \label{eq:conditions-ZN}
  1+a=N\alpha\ ,\qquad -3(1+a)=N\beta\ ,\qquad 3(1+a)=N\gamma\ ,
\end{equation}
with $\alpha,\beta,\gamma$ integers. Solutions to these
equations provide constraints between the $a$ charge and the cyclic
group order which when satisfied lead to inviable models, models where
the LO TBM structure does not hold due to hybridization, the remaining
choices are those for which an AF-inspired hybrid model becomes
possible. For $\beta\neq 0$ the system of equations in
(\ref{eq:conditions-ZN}) leads to
\begin{equation}
  \label{eq:N-a-values}
  N=-3\frac{1+a}{\beta}\ ,
\end{equation}
while for $\beta=0$ the solution corresponds to $a=-1$ for all
$N$. For a given negative (positive) value of $\beta$ there is a set
of positive (negative) values of $a$ which render $N$ integer, the
complete set of $(N,a)$ values thus determines the viability of the models. The resulting sets for
values of $\beta=-1, -2, -3, -4$ are displayed in
Fig. \ref{fig:n-and-a-assignments-allowing-dangerous-T}. The colored
dots indicate those $(N,a)$ choices for which all or some of the
dangerous operators will be present, while the remaining pairs
correspond to viable $(N,a)$ choices, apart from $N=3$ for which no
viable models can be considered no matter the values of the $a$
charge.

\subsection{The type-II seesaw case}
\label{sec:type-II-seesaw-case}
In the absence of further degrees of freedom, the general type-II
superpotential involves the following terms\footnote{For a
  contribution to the full neutrino mass matrix a single triplet
  suffices, however in the supersymmetric case two triplets with
  opposite hypercharge are needed to avoid the triangle gauge
  anomaly.}
\begin{equation}
  \label{eq:superpotential-type-II}
  W^{(II)}_\nu=y_\Delta LL \Delta_u + \lambda_u H_d H_d \Delta_u 
  + \lambda_d H_u H_u \Delta_d + \mu_\Delta \Delta_u \Delta_d\ ,
\end{equation}
with $Y(\Delta_u)=+1$, $Y(\Delta_d)=-1$ and lepton number charges
assigned according to $L(\Delta_u)=-2$ and $L(\Delta_d)=+2$. Since we
allow only renormalizable terms,
$\xi'$ can only be coupled to the last term, in which case
$A_4$ invariance leads to three
possibilities \footnote{$\Delta_d$ must couple to
  $H_u$ so the electroweak triplets can not be $A_4$ triplets.}
\begin{align}
  \label{eq:possibilities-type-II}
  W^{(II)}_\nu\supset
  \left\{
  \begin{aligned}
    &{\boldsymbol{(A)}}:\hspace{0.3cm}
    \lambda_\Delta\,\Delta_u'\,\Delta_d'\,\xi': A_4\;\;\text{non-trivial}\;\;\Delta_u\ , \Delta_d\ , \\
    &{\boldsymbol{(B)}}:\hspace{0.3cm}
    \lambda_\Delta\,\Delta_u''\,\Delta_d\,\xi': A_4\;\;\text{non-trivial}\;\;\Delta_u\ , \\
    &{\boldsymbol{(C)}}:\hspace{0.3cm}
    \lambda_\Delta\,\Delta_u\,\Delta_d''\,\xi': A_4\;\;\text{non-trivial}\;\;\Delta_d\ .
  \end{aligned}
  \right .
\end{align}
Bearing in mind that the new $\Delta L=2$ operator should have $\xi'$ appearing twice, general conclusions regarding these
possibilities can be drawn:
\begin{enumerate}
\item[$\boldsymbol{A.}$]
  Here, the electroweak triplet interactions reduces to:
  \begin{equation}
    \label{eq:A4-neutral-case-SP}
    W_\nu^{(II,A)}=y_\Delta\{LL\}''\Delta_u' + \lambda_\Delta\Delta_u'\Delta_d'\xi'\ ,
  \end{equation}
  which by itself does not allow the construction of a $\Delta L=2$
  operator. 
Adding a second triplet pair
however allows constructing
  such an operator, with  $\Delta_{u_1}''$ transforms as $\mathbf{1''}$,
  $\Delta_{d_1}$ trivially and the coupling $\{LL\}'\Delta_{u_1}''$
  absent (otherwise if present will induce a leading order
  $\{LL\}''\xi'H_u H_u$ operator). The latter follows from the following $\mathbb{Z}_N$ transformations:
  \begin{equation}
    \label{eq:deltadp-transform-case1}
    \Delta_{u_1}''\to \eta_N^{a-1} \Delta_{u_1}''\ ,
    \qquad
    \Delta_{d_1}\to \eta_N^{-2a} \Delta_{d_1}\ .
  \end{equation}
  in which case the full superpotential will read
  \begin{equation}
    \label{eq:full-SP-case1}
    W_\nu^{(II)}=W_\nu^{(II,A)} 
    + \mu_{\Delta_1}\Delta_{u_1}'' \Delta_d' 
    + \lambda_{\Delta_1}\Delta_{u_1}'' \Delta_{d_1} \xi' 
    + \lambda_d H_u H_u \Delta_{d_1}\ ,
  \end{equation}
  from which then the operator illustrated in Fig.
  \ref{fig:effective-operator-neutrino-masses}-($b$) can be
  generated.
 Note also from the $\{LL\}''\Delta_u'$ contraction that this construction leads to a $\Xi'$ model.
 
 \begin{figure}
   \centering
   \includegraphics[scale=0.7]{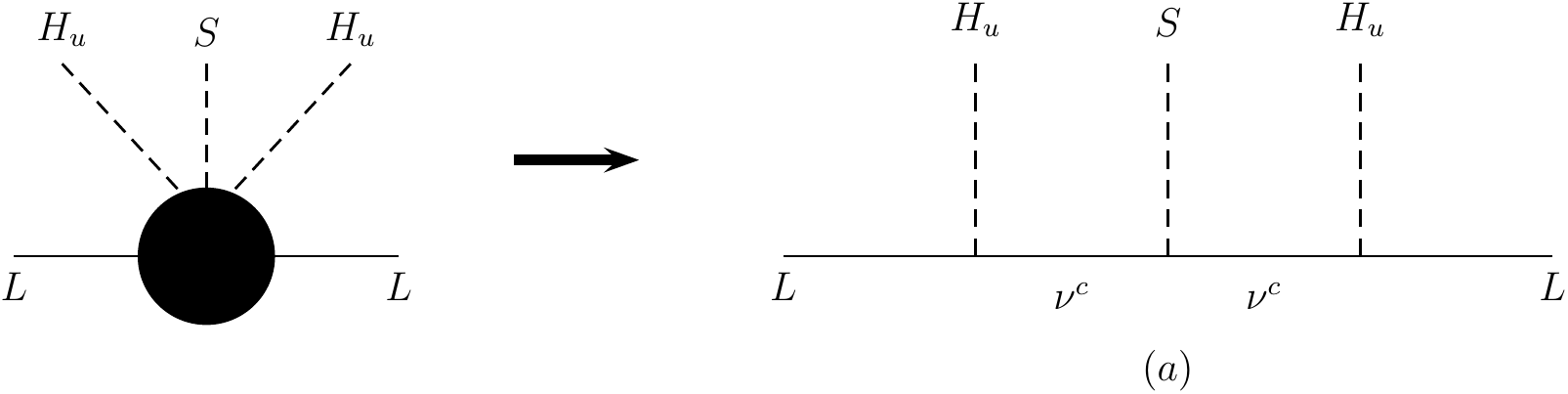}
   \vspace{0.7cm}\\
   \includegraphics[scale=0.7]{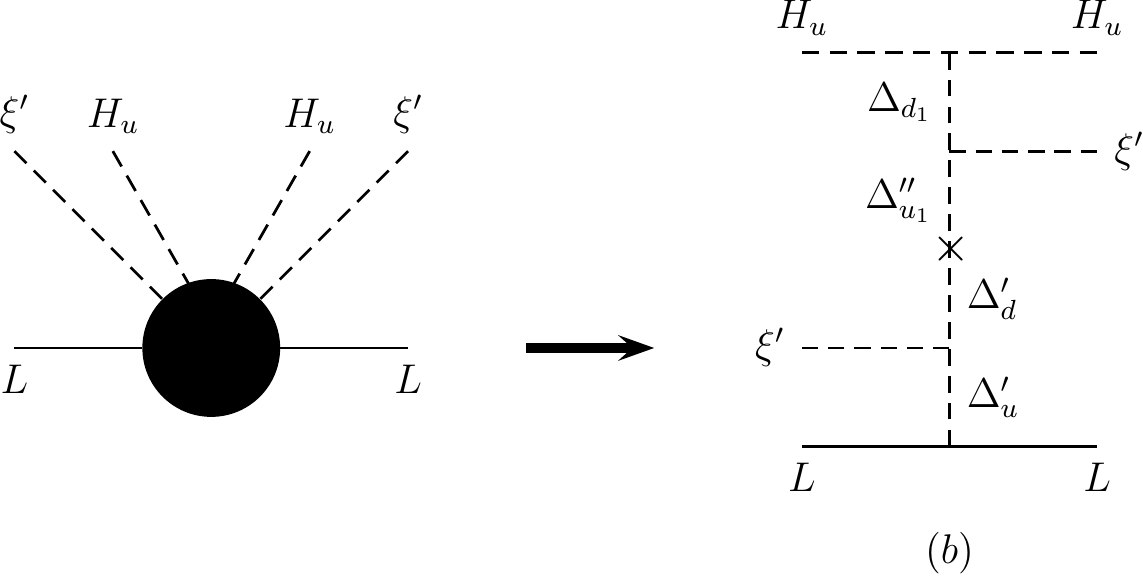}
   \caption{\it Lepton-number-breaking effective operators arising from
     integrating out the heavy type-I and type-II seesaw states, and
     responsible, after flavor and electroweak symmetry breaking, for
     the effective neutrino mass matrix with leading TBM-form (Feynman
     diagrams on top) and small deviations (Feynman diagrams on bottom)
     accounting for a non-zero reactor mixing angle. For the diagrams
     on top, $S$ stands for $\xi$, $\tilde \xi$ and $\phi_\nu$.}
   \label{fig:effective-operator-neutrino-masses}
 \end{figure}
\item[$\boldsymbol{B.}$]
 In this case the $A_4$
  transformation properties of the electroweak triplets allow the
  superpotential to have an extra term, namely
    \begin{equation}
    \label{eq:A4-charged-Du}
    W_\nu^{(II,B)}=y_\Delta\{LL\}'\Delta_u'' + \lambda_\Delta\Delta_u''\Delta_d\xi'
    +\lambda_d H_u H_u \Delta_d\ .
  \end{equation}
  This superpotential allows the construction of a $\Delta L=2$
  operator $\{LL\}'$ with a single $\xi'$, where the charge of $\xi'$ would be $+2+a$, leading to the problematic terms in (\ref{eq:new-driving-SP}).
So, assuring hybridization in this case
  requires the $\mathbb{Z}_N$ symmetry to forbid either the first or
  the last terms in the superpotential in (\ref{eq:A4-charged-Du})
  while keeping the remaining.  The first term is forbidden provided
  the triplets satisfy:
  \begin{equation}
    \label{eq:case2-LLabsent-transforms}
    \Delta_d\to \eta_N^{-2a} \Delta_d\ ,
    \qquad
    \Delta_u''\to \eta_N^{a-1} \Delta_u''\ ,
  \end{equation}
  whereas the last term will be absent as long as the following
  transformations hold
  \begin{equation}
    \label{eq:case2-DHuHuabsent-transforms}
    \Delta_u''\to \eta_N^{-2} \Delta_u''\ ,
    \qquad
    \Delta_d\to \eta_N^{a-1} \Delta_d\ ,
  \end{equation}
  We want to generate a $\Delta L=2$ operator with two $\xi'$, and in the case where the last term in
  (\ref{eq:A4-charged-Du}) is absent this requires three additional triplet
  pairs, so we will not consider this possibility in further detail. Instead,
on the other option the extra fields
  enabling the construction of the desirable operator correspond to a
  pair of triplets transforming as a $A_4$ singlets
  $\mathbf{1}'$, i.e. the same field content as the previous case. Indeed, renaming the fields so that the $A_4$ representations match the previous nomenclature we conclude that the $\mathbb{Z}_N$ charges are unchanged, and thus the superpotential turns out to be given exactly by (\ref{eq:full-SP-case1}), leading to Fig.
  \ref{fig:effective-operator-neutrino-masses}-($b$) and the same $\Xi'$ model.
\item[$\boldsymbol{C.}$] 
In this case the superpotential
  involves instead the coupling $H_d H_d \Delta_u$, thus it reads
  \begin{equation}
    \label{eq:superpotential-Deltau-neutral-case}
    W_\nu^{(II,C)}=y_\Delta\{LL\}\Delta_u + \lambda_\Delta\Delta_u\Delta_d''\xi'
    +\lambda_d H_d H_d \Delta_u\ .
  \end{equation} 
The dimension seven
  operator can not be constructed, and in contrast to the previous two
  cases, adding messengers does not fix the problem. This can be
  readily understood by noting that the requirement of having the $H_u
  H_u \Delta_d$ coupling calls for the extra scalar triplets involving
  a ``down-type'' triplet transforming trivially under $A_4$,
  $\Delta_{d_1}$, which in turn requires the presence of a ``up-type''
  triplet, $\Delta_{u_1}$, transforming trivially under $A_4$ too. The
  presence of both $\Delta_{u_1}$ and $\Delta_{d_1}$ lead to new terms
  which---via the $\mathbb{Z}_N$ symmetry---cannot be forbidden
  simultaneously. Thus, in this case a leading order dimension five
  operator will be unavoidable. It might be that by allowing extra
  triplet pairs the dimension seven operator can be constructed in the
  absence of a dimension five operator, here however we will not add
  further comments on this possibility.
\end{enumerate}
The full models can then be outlined as follows: the type-I seesaw
sector corresponds to a UV complete AF-inspired model as in \cite{Varzielas:2010mp}
and so at the LO produces a TBM flavor structure (without additional NLO contributions due to the specific UV completion). The sector dictating the
hybridization is a type-II seesaw
sector  with a flavor structure driven by the same $A_4$
symmetry, with a non-trivial flavon singlet responsible for the deviations, that only couples through the electroweak triplets.
Although at the renormalizable level the singlet
can be coupled to the electroweak triplets in three different manners,
only two choices allow for minimality (in the sense of the number of
triplet pairs)
and they both lead to the same model. Determined by the {\it hybridization statement},
minimality requires the type-II sector to involve two triplet
pairs. Tab. \ref{tab:messengers-type-I-type-II} shows the
transformation properties of the different messenger fields involved
in the minimal $A_4$-based hybrid $\Xi'$ model with type-I and type-II seesaw.
The $\Xi''$ model can be obtained by swapping $\xi'$, $\xi^{0'}$ with $\xi''$, $\xi^{0''}$ and also adjusting the messengers accordingly.

\begin{table}
  \centering
  \begin{tabular}{|c||c||cccc||cc|}
    \hline
    Fields &
    $\nu^c$ & 
    $\Delta_u'$ & $\Delta_d'$ & $\Delta_{u_1}''$ & $\Delta_{d_1}$ &
    $\xi'$ & $\xi^{0'}$
    \\\hline\hline
    $A_4$ & $\mathbf{3}$ & $\mathbf{1}'$ & $\mathbf{1}'$ & $\mathbf{1}''$ & 
    $\mathbf{1}$ 
    & $\mathbf{1}'$ & $\mathbf{1}'$\\
    $\mathbb{Z}_N$ &$-1-a$ & $-2$ & $1-a$ & $-1+a$ & $-2a$
    & $1+a$ & $-2(1+a)$\\
    $U(1)_\text{FN}$ & 0 & 0 & 0 & 0 & 0 & 0 & 0\\
    $U(1)_R$ & 0 & 0 & 2 & 0 & 2 & 0 & 2\\
    $U(1)_Y$ & $0$ & $+1$ & $-1$ & $+1$ & $-1$ & $0$ & $0$\\\hline
  \end{tabular}
  \caption{\it Transformation properties of the messenger 
    fields in the type-I and type-II sectors of the 
    $A_4$-based hybrid model. The last two columns correspond 
    to the flavon fields needed in the construction 
    of an $A_4$-based hybrid model.  \label{tab:messengers-type-I-type-II}}
\end{table}

We finally come to the issue of the singlet $\xi'$ vev. Once getting
rid of the dangerous terms in
(\ref{eq:superpotential-driving-sector-model}) the new ``driving''
sector superpotential reduces to:
\begin{equation}
  \label{eq:sp-driving-definitive}
  W_\phi^\text{New}=\bar g_4 \xi^{0'}\{\phi_l\phi_\nu\}''
  + \bar g_5 \xi^{0'}\xi'\xi'\ ,
\end{equation}
from which the minimization condition $\partial
W_\phi^\text{New}/\partial \xi^{0'}=0$ allows the determination of the
$\xi'$ vev, namely
\begin{equation}
  \label{eq:xiprime-vev}
  \langle\xi'\rangle\equiv \bar u
  =\sqrt{-\frac{\bar g_4\;v_l\;v_\nu}{\bar g_5}} \ .
\end{equation}

\subsection{The type-III seesaw case} 
\label{sec:type-iii}
\begin{figure}
  \centering
  \includegraphics[scale=0.8]{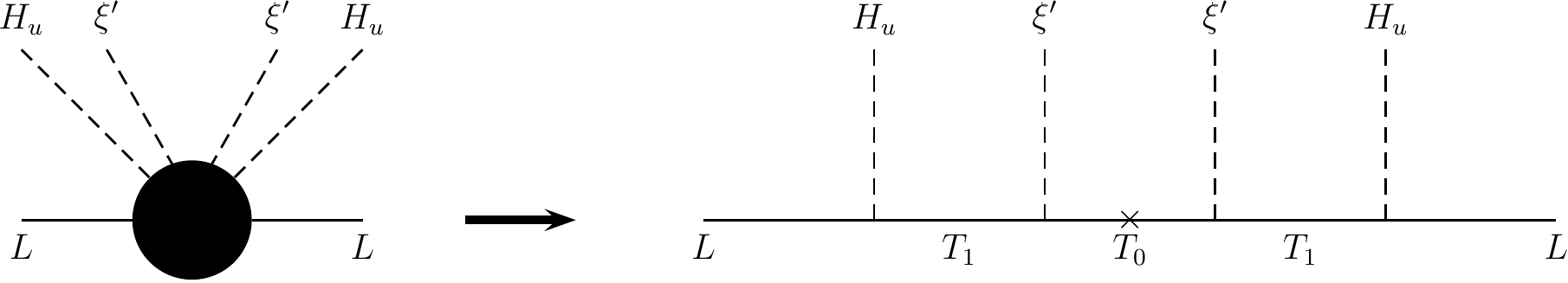}
  \caption{\it Lepton-number-breaking effective operator arising from
    integrating out the heavy type-III seesaw states.}
  \label{fig:effective-operator-neutrino-masses-iii}
\end{figure}

We can replace type-II seesaw with type-III seesaw to obtain different
UV completions. As it turns out, within this case there is a minimal
model requiring two additional superfields (compared to four in
Sec. \ref{sec:type-II-seesaw-case}).  We add superfields $T_0$ and
$T_1$:
\begin{equation}
  \label{eq:superpotential-type-III}
  W^{(III)}_\nu=y_T \{L\,T_1\}H_u  + \{T_1\,T_1\}\xi + \{\phi_\nu T_1 T_1\} +  \lambda_T \{T_1\,T_0\}''\xi' 
  + \mu_T \{T_0 \,T_0\} ,
\end{equation}
The assignments are quite distinct from the type-II case: here the
messengers are $A_4$ triplets with R-charge 1 like $\nu^c$, as shown
in see Tab. \ref{tab:messengers-type-III}.
\begin{table}
  \centering
  \begin{tabular}{|c||c||cc||cc|}
    \hline
    Fields &
    $\nu^c$ & 
    $T_1$ & $T_0$ & $\xi'$ & $\xi^{0'}$
    \\\hline\hline
    $A_4$ & $\mathbf{3}$ & $\mathbf{3}$ & $\mathbf{3}$ & $\mathbf{1}'$ & $\mathbf{1}'$\\
    $\mathbb{Z}_N$ &$-1-a$ & $-1-a$ & $0$ & $1+a$ & $-2(1+a)$\\
    $U(1)_\text{FN}$ & 0 & 0 & 0 & 0 & 0\\
    $U(1)_R$ & 1 & 1 & 1 & 0 & 2\\
    $U(1)_Y$ & $0$ & $0$ & $0$ & $0$ & $0$\\\hline
  \end{tabular}
  \caption{\it Transformation properties of the messenger 
    fields in the type-I and type-III sectors of the 
    $A_4$-based hybrid model.
      \label{tab:messengers-type-III}}
\end{table}
As $T_1$ and $\nu^c$ share
all the assignments (apart from $SU(2)$), the terms
$y_T \{L\,T_1\}H_u$, $\{T_1\,T_1\}\xi$ and $\{\phi_\nu T_1 T_1\}$ appear. The eigenvectors \cite{Sierra:2013ypa} of mass matrix generated by these operators alone 
are of TBM-form, so even though they arise from type-III they can be absorbed into $A$ and $B$
in (\ref{eq:Dirac-RHN-MM}). It is the presence of $T_0$ that distinguishes type-III from type-I. The vev of $\xi'$ can be obtained through (\ref{eq:xiprime-vev}) as in Sec. \ref{sec:type-II-seesaw-case}.
The diagram associated with the deviations is illustrated in
Fig. \ref{fig:effective-operator-neutrino-masses-iii}.  Checking the
$A_4$ triplet indices through diagram one can correctly identify
that this particular UV completion corresponds to a $\{L\,L\}''$
contraction, i.e. a $\Xi'$ model (the $\Xi''$ model corresponds to 
swapping $\xi'$, $\xi^{0'}$ with $\xi''$, $\xi^{0''}$).


\section{Conclusions}
\label{sec:concl}
Regardless of the mechanism responsible for neutrino masses, the
neutrino mass matrix should have a leading TBM-form with small
deviations accounting for a non-vanishing reactor mixing angle, as can
be proved---in a model-independent fashion---with the aid of the
eigenvector decomposition of the neutrino mass matrix combined with
current neutrino oscillation data \cite{Sierra:2013ypa}. Although such
structure can emerge in a variety of ways, here we take the position that
it suggests different mechanisms are participating
in the generation of neutrino masses, i.e. that we have ``hybrid neutrino masses''.

We have studied the feasibility of this ``interpretation'' under
rather general conditions, without specifying the mechanisms at
work. Our analysis then progressed by assuming an
underlying $A_4$ flavor symmetry in the lepton sector, and through
$\Delta L=2$ LO effective operators. Enforcing one of these operators
to generate a TBM-form structure, we investigated the
minimal $A_4$ representations capable of producing deviations from
TBM in agreement with data. After identifying these representations we
have proved that independently of the mechanisms assumed, fixing the reactor
angle in its experimental range always yield solar and atmospheric
mixing angles consistent with data. In this sense, deviations from TBM
induced by ``hybrid neutrino masses'' alleviate the problem found in
typical $A_4$ models, where NLO effective operators producing a
non-zero reactor angle quite often lead to values for the solar and
atmospheric angles which are inconsistent with data
\cite{Lin:2009bw, Altarelli:2012bn, Varzielas:2012ai}.

In order to illustrate this picture, we have constructed an UV
completion based on an interplay between type-I and either type-II or type-III
seesaw. Still under the assumption of an $A_4$ flavor symmetry, the
type-I seesaw contribution has been taken as responsible for the
leading TBM-form, while the deviations are driven by the other
contribution.
\section{Acknowledgments}
DAS wants to thanks Avelino Vicente for useful conversations, also
NORDITA for hospitality during the completion of this work.  DAS is
supported by a ``Charg\'e de Recherches'' contract funded by the
Belgian FNRS agency.  The work of IdMV is supported by the Swiss
National Science Foundation.


\end{document}